\documentclass{article}
\usepackage{spconf,amsmath,graphicx}
\usepackage{amsfonts}
\usepackage{bm}
\usepackage{wrapfig}
\usepackage{booktabs}

\title{TridentSE: Guiding Speech Enhancement with 32 Global Tokens}
\name{Dacheng Yin$^{1*}$\thanks{*Work done during internship at Microsoft Research Asia.}, Zhiyuan Zhao$^{2}$, Chuanxin Tang$^{2}$, Zhiwei Xiong$^{1}$, Chong Luo$^{2}$ }
\address{ $^{1}$University of Science and Technology of China, Hefei, China \\
        $^{2}$Microsoft Research Asia, Beijing, China
}
\begin{document}
\ninept
\maketitle
\begin{abstract}
In this paper, we present TridentSE, a novel architecture for speech enhancement, which is capable of efficiently capturing both global information and local details. TridentSE maintains T-F bin level representation to capture details, and uses a small number of global tokens to process the global information. Information is propagated between the local and the global representations through cross attention modules. To capture both inter- and intra-frame information, the global tokens are divided into two groups to process along the time and the frequency axis respectively. A metric discriminator is further employed to guide our model to achieve higher perceptual quality. Even with significantly lower computational cost, TridentSE outperforms a variety of previous speech enhancement methods, achieving a PESQ of 3.47 on VoiceBank+DEMAND dataset and a PESQ of 3.44 on DNS no-reverb test set. Visualization shows that the global tokens learn diverse and interpretable global patterns.
\end{abstract}
\begin{keywords}
Speech enhancement, global representation
\end{keywords}
\section{Introduction}
\label{sec:intro}

Speech enhancement (SE) aims to improve the quality of speech when it is contaminated with noise. In the deep learning era, speech enhancement techniques have also made great progress. One line of research is the time-domain methods \cite{convtasnet,demucs,se-conformer}, which process speech directly in waveform domain. Another line of research is the frequency-domain methods \cite{phasen,dccrn,dptfsnet}, which process speech in the T-F spectrogram domain. Our method belongs to the second category, and the objective of this research is to design an effective frequency-domain method for single-channel speech enhancement. 

For a frequency-domain SE method, the input is the time-frequency (T-F) spectrogram and previous research \cite{zhao2017two} has shown that it is better to use T-F mask instead of the T-F values as the immediate prediction target. Therefore, SE solves a dense classification or prediction problem, where dense means each T-F bin has a corresponding prediction output. The T-F bin level details, especially the phase structure, has become increasingly important with the development of masking method \cite{ibm, irm, psm, cirm}. This requires the SE network to faithfully capture the local details. On the other hand, previous work indicate that a good SE network is inseparable from the understanding of global (long-range) information on both frequency axis \cite{phasen} and time axis \cite{tftnet}. In short, an SE network needs to learn both local details and global information. 

Simultaneous learning of these two types of information is a non-trivial problem. Existing frequency-domain SE methods either adopt a cylindrical network structure \cite{ phasen, dptfsnet, tftnet} or a U-shaped structure \cite{uformer,dccrn}. In the first category, the feature map maintains its original T-F resolution as it is transformed by the SE network. While dense local information is naturally processed in each T-F bin, sparse global information is also aggregated by each T-F bin without much coordination, which is computationally inefficient.
In the second category, the feature map is gradually down-sampled during feature transformation. At its smallest size, it is affordable to compute the global semantic information. Then, the transformed feature map is gradually up-sampled to its original size. Skip connections are adopted to connect two layers with the same feature size before and after the computation on the smallest feature map, or in other words, merge the low-level and high-level features. Specially, the full resolution feature is not merged with high-level features until the end of the network. This limits the network's information fusion ability compared to the cylindrical architecture which process both level information at each layer.

In this work, we propose a third network structure for the SE task. It is formed by a main network, which maintains a full-resolution feature map, and two companion branches, each of which only keeps 16 global tokens. 
As the three-branch network architecture is like a trident, we name our SE network TridentSE.
The main network is responsible for computing the dense low-level details and the companion branches handle the global information. We differentiate temporal tokens and frequency tokens, both of which are initially extracted from the original feature map by the cross-attention operation. After each processing unit, they inject the global temporal and frequency information back to the main network by the same cross-attention operation. 

Employing two dedicated branches to compute high-level semantic information brings notable benefits. Compared with the cylindrical network structure, TridentSE significantly reduces the computational redundancy of global information. Compared with the U-shaped network structure, TridentSE is able to perform long-range computation from the very beginning of the network. The cross-attention-based fusion is also more powerful than the simple addition operation usually adopted in skip connection. 

Experimental results show that TridentSE achieves higher enhancement quality with lower computational complexity compared with various previous methods.
Through visualization, we confirm that the global tokens learn diverse and interpretable global patterns.

\begin{figure*}[t]
\centering
\includegraphics[width=\linewidth]{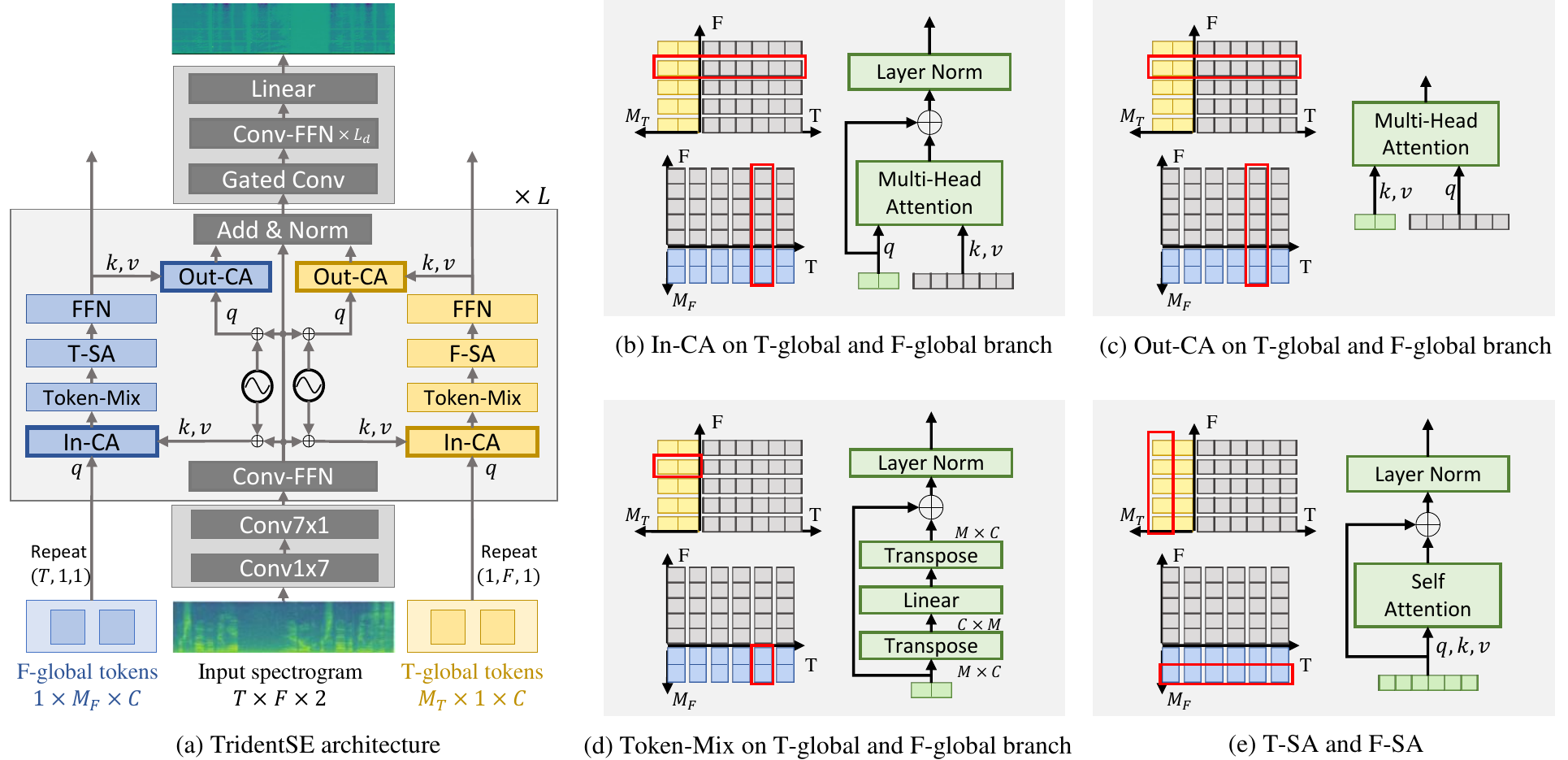}
\vspace{-1.5em}
\caption{\looseness -1 Architecture of the proposed TridentSE. \label{fig:arch}}
\vspace{-1.5em}
\end{figure*}

\section{Method}
\subsection{Overview of TridentSE}

We adopt the T-F masking framework in \cite{tftnet} to estimate the clean waveform from input noisy signal.
Fig.\ref{fig:arch} shows the overall architecture of the proposed T-F mask prediction network TridentSE. 
It is composed of three components: the encoder, the backbone, and the decoder. The encoder extracts local time-frequency feature from the input spectrogram, and the decoder decodes the time-frequency representation into the complex ratio mask $\bm{M_c} \in \mathbb{R}^{T\times F\times 2}$. Specifically, the encoder consists of two convolutional blocks, both of which have $C$ output channels, and the kernel sizes of the two conv blocks are $1\times 7$ and $7\times 1$, respectively. Each convolution operation is followed by batch normalization (BN) and ReLU activation. The decoder consists of a $1\times 1$ gated-convolutional layer, $L_d$ depth-wise separable convolutional blocks (Conv-FFN), and a linear layer that maps the $C$-dimensional feature vectors into complex numbers. The amplitude of the final output is restricted by Tanh activation. Each conv layer in Conv-FFN is followed by GELU activation, and each Conv-FFN is equipped with residue connection and post layer normalization. The major part of our network is the backbone, which is a stack of $L$ trident blocks.

\subsection{Trident block}
Each trident block consists of three branches: one main branch $\mathbf{B}_{m}$ and two companion branches, namely, time-domain global (T-global) branch $\mathbf{B}_{t}$, and frequency-domain global (F-global) branch $\mathbf{B}_{f}$. $\mathbf{B}_{m}$ calculates local information for each T-F bin. This is achieved with a Conv-FFN module, which uses a 2D depth-wise separable conv to aggregate the local information within the range of kernel size $K\times K$.

For the two companion branches, 
$\mathbf{B}_{t}$ and $\mathbf{B}_{f}$ form a duality pair, switching the notation between the two branches can be achieved by simply changing the subscipt between $T$ (or $t$) and $F$ (or $f$). Therefore, we only describe the network architecture in $\mathbf{B}_{t}$ for simplicity. 
$\mathbf{B}_{t}$ maintains full frequency resolution while reduces token number on the time axis into $M_T$. 
The initial feature of this branch $X^{0}_t\in \mathbb{R}^{M_T \times F \times C}$ is obtained by repeating a bank of $M_T$ initial global tokens $G_t\in \mathbb{R}^{M_T \times 1\times C}$ for $F$ times along the frequency axis, where $G_t$ is a learnable model parameter.
The calculation of $\mathbf{B}_{t}$ involves three modules. Firstly, as shown in Fig.\ref{fig:arch} (d) and (e), the token-mix and the frequency self-attention (F-SA) module mix the feature vector along $M_T$ and $F$ axis, respectively. Then, the feed-forward network (FFN) transforms the feature along the channel dimension. 

The information communication between the main branch and the companion branches are achieved by input cross-attention (In-CA) and output-cross-attention (Out-CA) modules, marked by thick border in Fig. \ref{fig:arch} (a). As shown in Fig. \ref{fig:arch} (b) and (c), each row of T-global feature interacts with the same row of full-resolution feature and each column of F-global feature interacts with the same column of the full-resolution feature. We choose cross-attention (CA) as the information communication method because it enjoys the flexibility of transforming between features of different token numbers. In addition, it calculates dynamic weights for information aggregation and broadcast, which can adapt to the variation among different input mixture signals.

To let the information communication aware of the time-frequency structure, the main branch feature is concatenated with sinusoidal 2D-positional encoding before fed into In-CA and Out-CA. 
The sub-modules introduced above, including Conv-FFN, FFN, In-CA, Token-Mix, and T-SA are equipped with residual connection and post layer normalization. GELU activation is used in the hidden layers of both Conv-FFN and FFN.

In summary, the whole model is mainly characterized by the following 6 hyper-parameters: channel number $C$, convolution kernel size $K$, the number of global tokens $M_T, M_F$, the number of trident blocks $L$, and the number of decoder Conv-FFNs $L_d$.
\subsection{Loss function}
Our loss function is applied on both waveform domain and spectrogram domain. For the loss on spectrogram domain, we follow \cite{phasen}, which uses power-compressed amplitude MSE loss $L_a$ and phase-aware MSE loss $L_p$. In the waveform domain, we also use MSE to calculate loss $L_w$. To directly optimize the PESQ score, we follow the method in MetricGAN \cite{metricgan} which introduces a metric discriminator $D$ that predicts differentiable PESQ score for SE network training. This brings an additional loss term $L_{GAN}$ for the predicted spectrogram.
Our total loss $L$ is the combination of the four losses introduced above. In summary, the loss is calculated as follows:
\begin{align}
    &L_a = MSE(|\hat{\bm{S}}|^{p}, |\bm{S}|^{p}),  \\
    &L_p = MSE(\hat{\bm{S}}/|\hat{\bm{S}}|^{1-p}, \bm{S}/|\bm{S}|^{1-p}),  \\
    &L_w = MSE(\hat{\bm{s}}, \bm{s}), \\
    &L_{GAN} = ||1 - D(\bm{S}, \hat{\bm{S}})||^2\\
    &L = (L_a + L_p + L_w) / 3 + \lambda L_{GAN},
\end{align}
where $\hat{\bm{S}}$, $\hat{\bm{s}}$, $\bm{S}$ and $\bm{s}$ are enhanced spectrogram and waveform, ground-truth clean spectrogram, and waveform respectively, $|\cdot|$ calculates the amplitude of the complex spectrogram, and $p$ is the power of the spectrogram compression. Here, we choose $p=0.3$. $\lambda$ is the GAN loss weight and we choose $\lambda=0.005$ in our experiment.

The loss for training the discriminator $D$ is calculated as follows:
\begin{align}
    &L_{D} = ||1 - D(\bm{S}, \bm{S})||^2 + ||Q(\bm{S}, \hat{\bm{S}}) - D(\bm{S}, \hat{\bm{S}})||^2,
\end{align}
where $Q(\bm{S}, \hat{\bm{S}})$ is the normalized PESQ score between $\bm{S}$ and $\hat{\bm{S}}$ ranged from 0 to 1.

\section{Experiments}
\subsection{Dataset and evaluation metrics} \label{exp:dataset}
We use two datasets to evaluate our method. The first one is the widely used VoiceBank+DEMAND dataset \cite{vcbkdemand} which contains paired clean and pre-mixed noisy speech. The clean speech samples are selected from the VoiceBank corpus \cite{vcbk}, where the training and test sets consist of 11,572 utterances from 28 speakers and 872 utterances from two speakers, respectively. For the noisy speech, the utterances in the training set are mixed with 10 types of noise (eight from DEMAND database and two artificially generated) at SNRs of 0, 5, 10, and 15 dB. In the test set, the utterances are mixed with five types of noise from the DEMAND database at SNRs of 2.5, 7.5, 12.5, and 17.5 dB. All the noise conditions and speakers in the test set are unseen in the training set. 

The second dataset is the large-scale DNS dataset \cite{dns}, which contains 500 hours of clean speech from 2150 speakers and over 180 hours of noise waveform from 150 classes. We perform online mixing during training stage to get noisy-clean pairs, where 75\% of the clean speech is convolved with randomly selected room impulse responses (RIR) provided in \cite{rir}, and the clean or reverberant speech is mixed with randomly selected noise with a uniformly sampled SNR ranged from -5 to 20 dB. The evaluation is done on two test sets named no\_reverb and with\_reverb, both of which contain 150 noisy-clean pairs.

We use a total of five metrics to evaluate the enhancement quality, all of which are better if higher. For both datasets, wide-band PESQ and short term objective intelligibility (STOI) are used to evaluate perceptual quality and intelligibility, respectively. In VoiceBank+DEMAND dataset, we use three additional mean opinion score (MOS) based metrics \cite{csigcbakcovl}: MOS prediction of the signal distortion (CSIG), MOS prediction of the intrusiveness of background noise (CBAK), and MOS prediction of the overall effect (COVL). All these three metrics are ranged from 1 to 5. 
For computational complexity evaluation, we report FLOPS for a 3-second input signal and the real time factor (RTF) on six Intel(R) Xeon(R) E5-2690 v3 CPU cores. We also report the model size in numbers of parameters.

\subsection{Implementation details} \label{exp:implementation}
All the utterances are resampled to 16kHz and we use 3-second segments for training.
STFT is computed using a Hann window of length 20ms, hop length of 10ms, and FFT size of 324. 
Four hyper-parameters, $M_T, M_F, L, L_d$, are tuned in our experiments. The other hyper-parameters are set as follows: $C=96$, $K=7$. The head number of T-SA, F-SA, In-CA, and Out-CA are set to 2, 2, 3, and 3, respectively. The hidden size of FFN and Conv-FFN is 96. The sinusoidal 2D-positional encoding has 64 channels.
The model is trained using LAMB \cite{lamb} optimizer with learning rate of 0.0008. The metric discriminator is trained with Adam \cite{adam} optimizer with a learning rate of 0.0004. The warm-up steps and batch size are set to 5,000 and 8, respectively. The training epochs are 300 and 120 for VoiceBank+DEMAND dataset and DNS dataset, respectively. 

\begin{table}[t]
\centering 
\vspace{-0.8em}
\caption{The effect of different configurations in TridentSE. \label{table:lat_alloc}}
\setlength{\tabcolsep}{1.0mm}
\vspace{-0.0em}
\begin{tabular}{ccccccccc}

\toprule
\# &$M_T$    & $M_F$    &MGAN    &$L$    &$L_D$    & FLOPS     & RTF      & PESQ      \\  \midrule
G1 &0        & 0        & w/o    & 6     & 6       & 18.3G     & 0.23     & 3.01      \\  
G2 &1        & 1        & w/o    & 3     & 4       & 23.6G     & 0.22     & 3.23      \\
G3 &2        & 2        & w/o    & 3     & 4       & 24.0G     & 0.22     & 3.24      \\
G4 &6        & 6        & w/o    & 3     & 4       & 25.3G     & 0.23     & 3.30      \\
G5 &16       & 16       & w/o    & 3     & 4       & 28.7G     & 0.24     & 3.31      \\ \midrule
M  &16       & 16       & w/     & 3     & 4       & 28.7G     & 0.24     & 3.44      \\  \midrule
A1 &axial    & axial    & w/     & 3     & 4       & 36.4G     & 0.35     & 3.41      \\   
A2 &1-group  & 1-group  & w/     & 3     & 4       & 18.9G     & 0.20     & 3.06      \\   \bottomrule
\end{tabular}
\vspace{-1.0em}
\end{table}

\begin{table*}[t]
\centering 
\caption{System comparison on VoiceBank+DEMAND dataset. Data with label '*' is our reproduced result. \label{table:sys_cpr}}
\vspace{-0em}
\begin{tabular}{llccccccccc}

\toprule
                         & Architecture   & PESQ   & CSIG  & CBAK  & COVL  & STOI(\%)& FLOPS & RTF    & \#Param.      \\  \midrule
Noisy                            & -      & 1.97   & 3.35  & 2.44  & 2.63  & 92.1    & 0     & 0        & 0         \\  \midrule
SEGAN \cite{segan}               &U-shaped& 2.16   & 3.48  & 2.94  & 2.80  & -       & -     & -        & -         \\  
DEMUCS \cite{demucs}             &U-shaped& 3.07   & 4.31  & 3.40  & 3.63  & 95      & 77.8G & 1.18     & 60.8M     \\ 
sudo-rm-rf \cite{sudo}           &U-shaped& 3.11*  & 4.36* & 3.58* & 3.74* & 95      & 21.9G & 0.20     & 4.85M     \\ 
SE-Conformer \cite{se-conformer} &U-shaped& 3.13   & 4.45  & 3.55  & 3.82  & 95      & -     & -        & -          \\  
\midrule
DCCRN \cite{dccrn}               &U-shaped& 2.68   & 3.88  & 3.18  & 3.27  & 94      & 25.2G & 0.26     & 3.67M      \\
TFT-Net \cite{tftnet}            &cylindrical& 2.75   & 3.93  & 3.44  & 3.34  & -    & 295G  & 0.73     & 5.81M      \\
PHASEN \cite{phasen}             &cylindrical& 2.99   & 4.21  & 3.55  & 3.62  & -       & 206G  & 0.51     & 20.9M      \\  
SN-Net \cite{snnet}              &cylindrical& 3.12   & 4.39  & 3.60  & 3.77  & -     & -     & -        & -          \\
DB-AIAT \cite{dbaiat}            &cylindrical& 3.31   & 4.61  & 3.75  & 3.96  & 96    & 68.0G   & 3.81  & 2.81M      \\  
DPT-FSNET \cite{dptfsnet}        &cylindrical& 3.33   & 4.58  & 3.72  & 4.00  & 96    & 55.7G* & 1.12*     & \textbf{0.88}M      \\
CMGAN     \cite{cmgan}           &cylindrical& 3.41   & 4.63  & \textbf{3.94}  & \textbf{4.12}  & 96    & 116G   & 1.02      & 1.83M     \\ \midrule
TridentSE-S                      &Trident & 3.36   & 4.61  & 3.75  & 3.99  & \textbf{96}  & \textbf{19.8G}  & \textbf{0.16}       & 1.00M      \\  
TridentSE-M                      &Trident & \textbf{3.44}   & \textbf{4.65}  & 3.77  & 4.06  & \textbf{96}  & 28.7G  & 0.24       & 1.42M      \\ 
TridentSE-L                      &Trident & \textbf{3.47}   & \textbf{4.70}  & 3.81  & 4.10  & \textbf{96}  & 59.8G   & 0.49      & 3.03M      \\ \bottomrule
\end{tabular}
\end{table*}

\subsection{Global representation and adversarial training}
Table \ref{table:lat_alloc} shows the experimental results on VoiceBank+DEMAND dataset. In experiment G1-G5, we gradually increase the global token number until PESQ does not increase. Here, G1's companion branches are removed, therefore we compensate the computational cost by adding layers. The comparison between G1 and G2 shows that adding the companion branches is important, since even if there is only one global token, it can bring a significant PESQ improvement of 0.22. By increasing the global token number, we can still get a relatively large PESQ imporvement of 0.09. This gain is saturated at a relatively small global token number of 16, which confirms that processing global information does not need dense calculation. Experiment M produces our full model TridentSE by adding adversarial training over G5. The adversarial training brings a significant PESQ improvement of 0.13. 

\subsection{Ablation study}
In experiment A1 and A2 of Table \ref{table:lat_alloc}, we carry out ablation study on the global information processing method. In experiment A1, we replace the two companion branches with two axial attention blocks that calculates attention along time and frequency axis respectively. The FLOPS and RTF increase by 27\% and 46\% respectively, while PESQ is decreased by 0.03. This result shows that Trident architecture is more powerful and efficient than the traditional axial attention in speech enhancement task. In A2, instead of processing along T and F axes using two companion branches, only one companion branch is used, and the global tokens directly aggregate information from all the T-F bins in the full-resolution feature. As a result, we observe a huge PESQ drop of 0.38. This indicates that separately processing along T and F axis is necessary. In summary, the proposed Trident architecture is a better choice than other global information processing methods we evaluated.

\subsection{System comparison} \label{exp:syscpr}

In Table \ref{table:sys_cpr}, we compare our method with other time-domain and T-F domain methods with U-shaped and cylindrical architecture on VoiceBank+DEMAND dataset. TridentSE-M is the same model as the experiment M in Table \ref{table:lat_alloc}. TridentSE-S and TridentSE-L are small and large version of TridentSE, respectively. The only difference is the model depth. In TridentSE-S, $L=L_d=2$, while in TridentSE-L, $L=7$ and $L_d=8$. TridentSE-S has the smallest FLOPS and RTF among all the listed methods, but the enhancement quality outperforms all other methods except CMGAN and the COVL score of DPT-FSNET. Compared with CMGAN, TridentSE-M and -L achieve higher PESQ and CSIG with only one-forth and one-half of computational cost, respectively. In summary, Trident architecture is faster and better than the previous methods. Table \ref{table:dns} shows the results on DNS dataset. With half of the inference time, TridentSE-L achieves a new state-of-the-art on PESQ and outperforms DPT-FSNET by a large margin on no\_reverb testset .

\begin{table}[h]
\centering 
\caption{Results on DNS no\_reverb / with\_reverb testset. \label{table:dns}}
\vspace{-0.0em}
\begin{tabular}{cccc}

\toprule
                                &PESQ                & STOI(\%)         & RTF        \\  \midrule
Noisy                           & 1.58 / 1.82        & 91.52 / 86.62    & 0     \\  \midrule
PoCoNet\cite{poconet}           & 2.75 / 2.83        & - / -            & -     \\
FullSubNet\cite{fullsub}        & 2.78 / 2.97        & 96.11 / 92.62    & 0.39  \\
DPT-FSNet\cite{dptfsnet}        & 3.26 / 3.53        & 97.68 / 95.23    & 1.12 \\  \midrule
TridentSE-L                     & \textbf{3.44} / 3.50        & \textbf{97.86} / 95.22    & 0.49          \\  \bottomrule

\end{tabular}
\end{table}

\subsection{Attention visualization} \label{exp:attnvis}
\begin{figure}[h]
\centering
\includegraphics[width=\columnwidth]{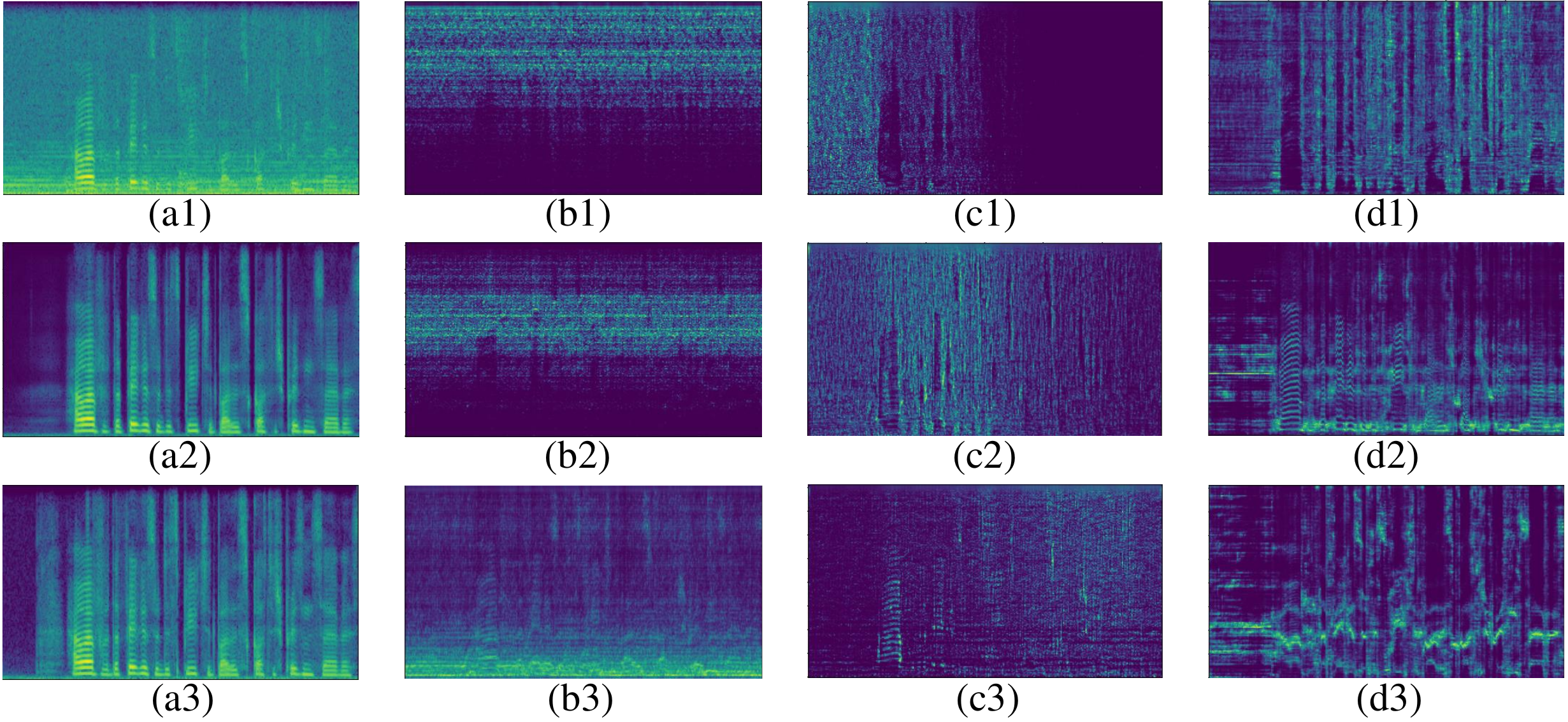}
\vspace{-1.5em}
\caption{\looseness -1 Visualization of spectrogram and attention maps. (a1-a3): STFT spectrogram of noisy, enhanced, and clean speech. (b1-b3 \& c1-c3): Shallow layer attention maps on F-global and T-global branch. (d1-d3): Attention maps that focus on speech or noise patterns \label{fig:attn_vis}}
\vspace{-1.5em}
\end{figure}

In this experiment, we figure out what is learned in the global tokens by visualizing In-CA's attention maps as shown in Figure \ref{fig:attn_vis}. 
The attention map is obtained by enhancing a sample with SNR of 1.4dB using TridentSE-L. 
(b1-b3) and (c1-c3) demonstrate the shallow layer attention of three global tokens in the frequency and time global branch, respectively. They mainly attend to different wide frequency bands and time spans. As the layer goes deeper, the attention map shows more speech-specific patterns, such as harmonics (d2) and formants (d3) which are the important evidence of identifying phonemes. Some other global tokens focus on noise-dominant T-F bins (d1) to capture noise-specific information. All these attention maps are distributed globally in the spectrogram. Therefore we can confirm that the global tokens have learned meaningful global information.

\section{Conclusion} \label{sec:conclusion}
We have presented a novel speech enhancement network named TridentSE. It adopts a trident network structure with a main network and two companion branches. The light-weight main network maintains full resolution of a spectrogram to capture low-level details in each T-F bin, while the companion branches use a total of 32 global tokens to efficiently process concentrated global information. Combined with adversarial training method, TridentSE achieves state-of-the-art performance on VoiceBank+DEMAND and DNS dataset with much lower computation than previous methods. The attention maps show that the global tokens have learned diverse and meaningful global information. 
In the future, we plan to design a causal version of TridentSE for low-delay real-time scenarios. 

\bibliographystyle{IEEEbib}
\bibliography{strings,refs}

\end{document}